\documentclass[english,12pt]{article}
\usepackage{color}
\usepackage{amssymb}
\usepackage{amsmath}
\usepackage{babel}
\usepackage{pifont}
\usepackage
[dvips,
colorlinks=true,
linkcolor=webgreen,
filecolor=webbrown,
citecolor=webgreen,
bookmarksopen=false,
]{hyperref}
\definecolor{webgreen}{rgb}{0,.5,0}
\definecolor{webbrown}{rgb}{.6,0,0}
\usepackage[T1]{fontenc}
\usepackage[cp1250]{inputenc}
\usepackage{array}
\usepackage[dvips]{graphicx}
\usepackage{amssymb}
\date{}
\definecolor{arcolor}{cmyk}{0.05,0.95,0.9,0.1}
\title{Quantum English Auctions}
\author{Edward W. Piotrowski\\ Institute of Theoretical Physics,
University of Bia\l ystok,\\ Lipowa 41, Pl 15424 Bia\l ystok,
Poland\\ e-mail: \href{mailto:ep@alpha.uwb.edu.pl}{ep@alpha.uwb.edu.pl}\\
 Jan S\l adkowski\\ Institute of Physics, University of Silesia, \\ Uniwersytecka
4, Pl 40007 Katowice, Poland \\ e-mail:
\href{mailto:sladk@us.edu.pl}{sladk@us.edu.pl}
}
\begin{document}
\maketitle
\def\Z{{\bf Z\!\!Z}}
\def\R{{\bf I\!R}}
\def\N{{\bf I\!N}}
\begin{abstract}
We continue the analysis of quantum-like description of markets
and economics. The approach has roots in the recently developed
quantum game theory and quantum computing. The present paper is
devoted to quantum English auction which are a special class of
quantum market games. The approach allows to calculate profit
intensities for various possible strategies.

\end{abstract}

 \vspace{5mm}

\section{Introduction} Recent research on quantum computation
and quantum information allowed to extend the scope game theory
for the quantum world \cite{1}-\cite{4}.  We showed how quantum
game theory may be used for describing financial market phenomena
\cite{5,6}. The purpose of this paper is to extent the previous
results to incorporate also quantum version of English auctions.
Such a generalization is desirable because auctions prevail among
market games and we think that quantum-like approach provide us
with more precise models of market phenomena than the standard
ones based on probability theory. The quantum-like description of
market phenomena has a remarkable chance of gaining favourable
reception from the experts. On the other hand only thorough
investigation may reveal if economics already is in or would ever
enter the domain of quantum theory. Quantum computation is on the
verge of being recognized as an autonomous scientific discipline
and  efforts to unify social and physical phenomena should not
cause astonishment \cite{7}. It might be that while observing the
due ceremonial of everyday market transaction we are in fact
observing capital flows resulting from quantum games eluding
classical description. " If human decisions can be traced to
microscopic quantum events one would expect that nature would have
taken advantage of quantum computation in evolving complex brains.
In that sense one could indeed say that quantum computers are
playing their market games according to quantum rules" \cite{11} .
\\ In the following sections we
consider quantum English auctions and analyze possible
profits gained under various conditions. Vickrey's auctions and
various generalizations would be presented in following papers.

\section{Quantum bargaining with one-side bidding} Let us consider
a particular case of quantum bargaining ($q$\/-bargaining) \cite{5,6}
in which the first player, denoted by -1 for future convenience,
sells a definite amount of some good and the second one, denoted
by $1$ want to buy the good in question. The player $1$ proposes a
price and the player -1 accept or reject the proposal. Their
polarizations \cite{6} are $|\mit0\rangle$ and $|\mit1\rangle$,
respectively so the $q$\/-bargaining has  the polarization
$|\mit0\rangle _{\text{-}1}|\mit1\rangle _{1}$. The transaction in
question is accomplished if the obvious rationality condition is
fulfilled
\begin{equation}
\label{haucja-war} [{\mathfrak q}+{\mathfrak p}\leq 0],
\end{equation}
where  the convenient Iverson notation \cite{8} is used
($[expression]$\/ denotes the logical value (1 or 0) of the
sentence $expression$) and the parameters ${\mathfrak p}=\ln
\mathfrak{c}_{\text{-}1}$\/ and $-{\mathfrak q}=\ln {\mathfrak
c}_1$\/ are random variables corresponding to prices at which the
respective players withdraw, the {\sl withdrawal prices}. The
random variables $\mathfrak{p}$ and $\mathfrak{q}$ describe
additively profits resulting from price variations. Their
probability densities are equal to squared absolute values of the
appropriate wave functions $\langle p|\psi\rangle_{\text{-}1}$\/
and $\langle q|\psi\rangle_1$ (that is their strategies).  Note
that the discussed $q$\/-bargaining may result from a situation where
several players have intention of buying but they were outbid by
the player $1$ (his withdrawal price ${\mathfrak c}_1$\/ was
greater than the  other players ones,
 ${\mathfrak c}_1>{\mathfrak c}_k$,
$k=2,\ldots,N$)\/. This means that all part in the auction are
fermions and they are subjected to the Pauli exclusion principle
according to which two players cannot occupy the same state. The
fermionic character of $q$\/-bargaining parts first noted in \cite{5}
in a slightly different context. If at the outset of the auction
there are several bidding players then the rationality condition
takes the form
\begin{equation}
\label{haucja-war2} [{\mathfrak q}_{\min} +{\mathfrak p}\leq 0]
\end{equation}
where ${\mathfrak q}_{\min}
:=\underset{k=1,\ldots,N}{\min}\{{\mathfrak q}_k\}$ is the
logarithm of the highest bid multiplied by -1. The probability
density of making the transaction with the $k$-th buyers at the
price $c_{k}={\mathrm e}^{-q_k}$\/ is according to Ref. [5] given
by
\begin{equation}
\label{haucja-dobicie}
dq_k \;\frac{|\langle q_k|\psi_k\rangle |^2}{
\langle\psi_k|\psi_k\rangle} \prod_{\substack{m=1\\ m\neq k}}^{N}
\int_{-\infty}^{\infty}\negthinspace\negthinspace\negthinspace dq_m \frac{|\langle q_m|\psi_m\rangle |^2
}{\langle\psi_m|\psi_m\rangle}\int_{-\infty}^{\infty}\negthinspace\negthinspace\negthinspace dp\;
\frac{|\langle p|\psi_{\text{-}1}\rangle |^2}{\langle\psi_{\text{-}1}|\psi_{\text{-}1}\rangle}
\;[\;q_k=\negthinspace\negthinspace\min_{n=1,\ldots,N}\{q_n\}\;]\;[q_k+p\leq0].
\end{equation}
The seller is not interested in making the deal with any
particular buyer and the unconditional probability of
accomplishing the transaction at the price $c$ is given by the sum
over $k\negthinspace=\negthinspace1,\ldots,N$\/ of the above formula
with $q_k\negthinspace=\negthinspace-\ln c$.
If we neglect the problem
of determining the probability amplitudes in
$(\ref{haucja-dobicie})$\/ we easily note that the discussed
$q$\/-bargaining is in fact the English auction (first price auction)
so popular on markets of rare goods. From the quantum context it
is interesting to note that the formula $(\ref{haucja-dobicie})$\
contains wave functions of payers who were outbid before the end
of the bargaining (cf the Pauli exclusion principle). The
probability density of "measuring" of a concrete value $q$ of the
random variable $\mathfrak{q}$ characterizing the player, according
to the probabilistic interpretation of quantum theory, is equal to the
squared absolute value of the normalized wave function describing
his strategy
\begin{equation}
\label{cica-eisert} \frac{|\langle
q|\psi_k\rangle|^2}{ \langle\psi_k|\psi_k\rangle}\,dq\;.
\end{equation}
Physicists normalize wave functions because conservation laws
require that. Therefore the trivial statement that if a market
player may be persuaded into making a deal or not is a matter of
price alone, corresponds to the physical fact that a particle
cannot vanish without any trace.
\section{Quantum English auction with a dominating bidder}  The
most frequent scenario of an English auction is the one with
public reserve price (bids lower than the reserve price are
rejected). The quantum version of such an action may defined as
the auction when measures are cut from one side and the player -1
does not fix his withdrawal price ($|\langle
q|\psi_{\text{-}1}\rangle |^2=\delta(q-q')$). We cannot identify
withdrawal and the reserve prices in quantum approach because this
would result in contradiction because it would entangle the
reserve price with the players -1' polarization which forbiden by
the Pauli exclusion principle (both players would wind up in the
same polarization
state before settlement of the bargain). \\
We restrict our analysis to quantum English auctions during which
the player -1 has fixed withdrawal price ${\mathfrak c}={\mathrm
e}^{p'}$. The corresponding probability measure is equal to
$|\langle p|\psi_{\text{-}1}\rangle |^2dp=\delta(p-p')dp$. We will
also suppose that players are allowed to use mixed strategies. In
that case the squared absolute values of probability amplitudes
$|\langle q_k|\psi\rangle |^2$ in $(\ref{haucja-dobicie})$ should
be replaced by appropriate convex linear combination
$\eta(q_k)$. \\
If for some $k\negthinspace=\negthinspace k'$\/ the formula
$(\ref{haucja-dobicie})$\/ might
by replaced by the measure
\begin{equation}
\label{haukcja-ryba}
[q_{k'}\negthinspace+p'\leq0]\;\eta(q_{k'}\negthinspace)\,dq_{k'}
\end{equation}
then the auction in question reduces the merchandising
mathematician model \cite{5,6} that is to  $q$\/-bargaining with the
polarization $|\mit0\rangle_{\text{-}1}|\mit1\rangle_1$ and the
player -1 strategy being a proper state of the operator of supply
$\mathcal{P}$ or operator of demand $\mathcal{Q}$ if she is
selling or buying, respectively. This may happen if the measure of
the set of events for which ${\mathfrak
q}_{k'}\neq\negthinspace\min_{n=1,\ldots,N}\{{\mathfrak
q}_n\negthinspace\}$ is negligible eg $k'$-th player offers are to
high for the rest of participants.
\section{Quantum English auction with identical strategies of
bidders} 
\begin{center}
 \begin{figure}[h]
 \begin{center}
\includegraphics[height=5.25cm, width=9.25cm]{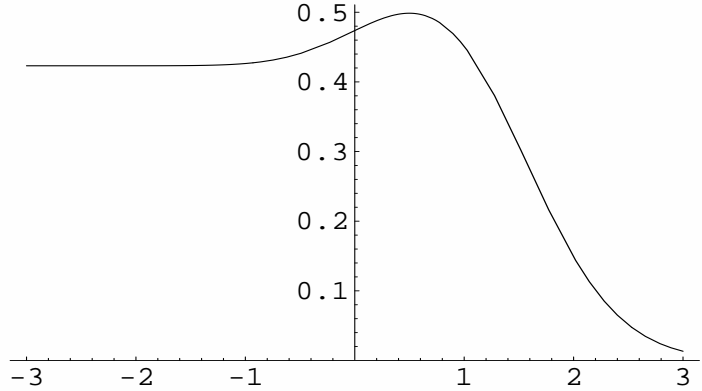}
\end{center}
\caption{Plot of profit intensity $\rho_N(p')$ in English
$q$\/-auction for {$N\negthinspace=\negthinspace 3$}}
 \label{trzy}
 \end{figure}
\end{center}

\begin{center}
 \begin{figure}[h]
 \begin{center}
\includegraphics[height=5.25cm, width=9.25cm]{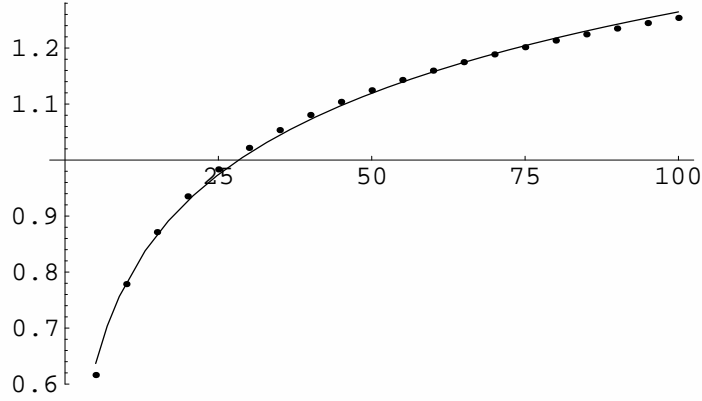}
\end{center}
\caption{Maximal values of profit intensities in English
$q$\/-auction for $N\negthinspace\leq\negthinspace 100$ against
the curve $0.21\log N + 0.3$}
 \label{fig:haukcjaluki}
 \end{figure}
\end{center}

Let us now consider the class of English auctions with
all $N$ buyers having the same density of distribution of the
logarithm of the withdrawal price,  $\eta(q)$\/, which may be
interpreted as the strategy of a equilibrium market with the mean
value of the withdrawal price equal to zero (one may always find
appropriate currency units). The formula
$(\ref{haucja-dobicie})$\/ reduces to
\begin{equation}
\label{haucja-dobicie1} [q+p'\leq0] \;\eta(q)\; \Bigl(
\int_{q}^{\infty} \negthinspace\negthinspace\eta(r)\; dr\;
\Bigr)^{N-1}dq
\end{equation}
because the probability of success in the auction with price
belonging to $[{\mathrm e}^{-q},{\mathrm e}^{-q}(1+dq)]$ does not
depend on the player. The random variable $-{\mathfrak q}$
represents the profits measured by the compound rate of return
achieved by the player -1 in the auction with respect to the
average market price of the good being sold. To measure the
profits of the seller is sufficient to notice that her situation
is identical to $q$\/-bargaining with fixed polarization. Her abstract
opponent being the Rest of the World \cite{5} might accomplish the
bargaining by bidding price whose logarithm with reversed sing is
a random variable ${\mathfrak q}'$ with the distribution equal to
$N$ times the distribution $(\ref{haucja-dobicie1})$ that is the
function ${\mathfrak q}':= \min\{{\mathfrak q}_1,\ldots,{\mathfrak
q}_N\}$ (1th-order statistics \cite{9}). The profit intensity of
the seller takes the form \cite{10}
\begin{equation}
\label{haukcja-ro} \rho_N(p'):={{E(-[{\mathfrak q}'+
p'\leq0]{\;\mathfrak q}') }\over{E({\mathfrak t}) }}=
          \frac{-\int\limits^{-p'}_{-\infty}q\;\eta(q)\;
        \bigl({\int\limits^{\infty}_{q}\eta(r)\; dr}\bigr)^{N-1}dq}
        {N^{-1} + \int\limits^{-p'}_{-\infty}\eta(q)\;
        \bigl({\int\limits^{\infty}_{q}\eta(r)\; dr}\bigr)^{N-1}dq}
\end{equation}
where $\mathfrak{t}$ is the random variable describing time needed
by the player -1 an average profit $E(-[{\mathfrak q}'+
p'\leq0]{\;\mathfrak q}')$ (with a fixed withdrawal price $p'$).
Let us recall that $(\ref{haukcja-ro})$ has a remarkable property
of attaining it maximal value at a fixed point, that, if
$\mathfrak q$ has normal distribution, is a contraction almost
everywhere. Normal distributions play a special role in quantum
market games models because they exhaust the class of positive
definite pure strategies \cite{5,6}. They  describe also
equilibrium markets. Therefore til the end of next paragrph we
will suppose that $\eta(q)$ is a normal distribution. If
$\rho_N(p')$ is a contraction then the opponent of bidders may use
a natural method of maximization of her profit intensity. The
method consists in repeated corrections of the withdrawal price up
to the value equal to mean profit intensity \cite{5,6,10}. The
knowledge of the character of the distribution $\eta(q)$ is not
necessary. But if the number of bidders is big the same result may
be achieved by setting the withdrawal price to zero
($p'\negthinspace=\negthinspace -\infty$). Fig. \ref{trzy}
presents the fall in values of the function $\rho_N(p')$ from
maximum to $\rho_N(-\infty)$.
The $N$-dependence become negligible for large values of $N$. Tab\mbox{.}
\ref{tab:ekla-haukcja} presents results gained while using two
methods of selection the strategy  of fixing withdrawal prices by
player -1 and $N\negthinspace\leq\negthinspace 10$.
The last column of Tab\mbox{.} \ref{tab:ekla-haukcja}
contains ratios that do not depend on the
dispersion $\sigma$ of $\eta(q)$.
\begin{table}
\begin{tabular}{rccc} \hline
  N & $\max_{\substack{p'}}\rho_N(p')$ & $\rho_N(-\infty)$&
  ${\max_{\substack{p'}}\rho_N(p')}/{\rho_N(-\infty)_{\hphantom{j}}}$
   \\\hline
  1 & 0.27603\phantom{1} & 0 & --\\
  2 & 0.410091 & 0.282095& 1.45373\\
 3 & 0.498606 &0.423142&1.17834  \\
  4 & 0.564273 & 0.514688& 1.09634\\
   5 & 0.616195 & 0.581482&1.0597\phantom{4} \\
    6 & 0.658949 &  0.633603& 1.04\phantom{234}\\
     7 & 0.695165 & 0.676089&1.02822 \\
      8 &0.726489  & 0.7118\phantom{10}&1.02064 \\
       9 & 0.754024 & 0.742507&1.01551 \\
 10 & 0.77854\phantom{1}  & 0.769376&1.01191 \\
   \hline
\end{tabular}
  \centering
  \caption{Profit intensities in units of $\sigma$)  Gaussian $1$th-order
  statistics}
  \label{tab:ekla-haukcja}
\end{table}
It is obvious that the attractiveness  of auction consists in not
in abilities of the seller but the rivalry among great number of
bidders. The opportunities resulting from growing number of
bidders present Fig\mbox{.} \ref{fig:haukcjaluki}.
It is easy to notice that for $N\negthinspace\leq\negthinspace100$
a very good approximation of the profits counted in units $\sigma$
is given by a logarithmic series if the assumption of equality of
Gaussian distributions $\eta(q_k)$ is valid. The player -1'
profits measured with respect to the mean value of the logarithm
of market price of the good being sold must be balanced by the
loss winning bidder (modulo the possible brokerage that we
neglect). It follows that in case the player -1 does not fix her
withdrawal price the intensity of average losses of  bidders is
equal to $-\frac{2\rho_N(-\infty)}{1+N}$. Therefore the increase
in the number of bidders is advantageous to both sides. It is not
possible to reduce the number of players by forming linear
combinations. Such a characteristics being a direct consequence of
the quantum no-delete theorem \cite{15} forbids manipulations on
the quantum level and stabilizes the equilibrium gained by
pure strategies of bidders
acting as anonymous Rest of the World \cite{6}.

\section{Profit intensities asymptotic behaviour}
\begin{center}
 \begin{figure}
 \begin{center}
\includegraphics[height=5.25cm, width=9.25cm]{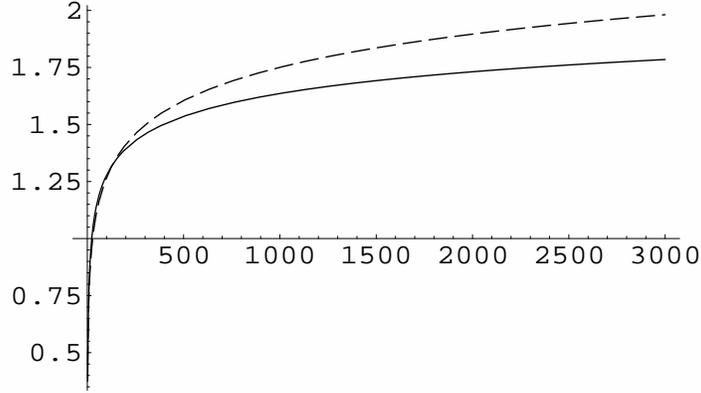}
\end{center}
\caption{Plots of fuctions $\sqrt{\frac{\ln
N}{2}}+\frac{2\gamma-\ln 4\pi-\ln\ln N}{4\sqrt{2\ln N}}$ and
$0.21\log N + 0.3$ (dashed line)} \label{fig:haukcjapodw}
\end{figure}
\end{center}

The present day growing popularity of internet auctions and almost
unlimited access to such auction organized by robots raises the of
maximal profit intensity in English $q$\/-auction with large
($N\negthinspace\gg\negthinspace100$) number of bidders. In this
case the approximation by the function $0.21\log N + 0.3$ is no
longer valid. But fortunately it is possible to find the
asymptotic behaviour of the function $\max_{p'}\{\rho_N(p')\}$. To
this end it is sufficient to find the asymptotic behaviour of the
random variable
\begin{equation}
\label{przeska-haukcja} a_N \mathfrak{q}'+b_N ,
\end{equation} where the series $a_N$ and $b_N$ are given by
$$
a_N:=\sqrt{2\ln N} \text{~~~~and~~~~} b_N:=\tfrac{1}{2}(\ln 4\pi+
\ln\ln N)-2\ln N .
$$
If $\eta(q)$ is the standard normal distribution then the
cumulative distribution function of the random variable
$(\ref{przeska-haukcja})$ being the rescaled logarithm of the
price striking the bargain tends to the Gumbel cumulative
distribution function (double exponential) \cite{12}
\cite{encyklopediamat}
$$
P\Bigl(a_n \mathfrak{q}'+b_n\leq
x\Bigr)\xrightarrow{N\rightarrow\infty} \text{e}^{-\text{e}^{-x}}
.
$$
The expectation value of a random variable with probability
density $\text{e}^{-\text{e}^{-x}-x}dx$ is equal to  the Euler
constant
$\gamma:=\lim_{N\rightarrow\infty}\sum_{k=1}^N\tfrac{1}{k}-\ln
N\simeq 0.5772$. This after same elementary algebra leads to the
asymptotic behaviour of the profit intensity
$\max_{p'}\{\rho_N(p')\negthinspace\}$ for
$N\negthinspace\rightarrow\negthinspace\infty$ of the form
$$
\sqrt{\frac{\ln N}{2}}+\frac{2\gamma-\ln 4\pi-\ln\ln
N}{4\sqrt{2\ln N}}.
$$
The difference between the above function and the previous
logarithmic approximation is plotted in Fig.
\ref{fig:haukcjapodw}.
Details may be found in Cram\'er's book \cite{9}.
\section{Bidder's profits} 
\begin{center}
 \begin{figure}[h]
 \begin{center}
\includegraphics[height=5.25cm, width=9.25cm]{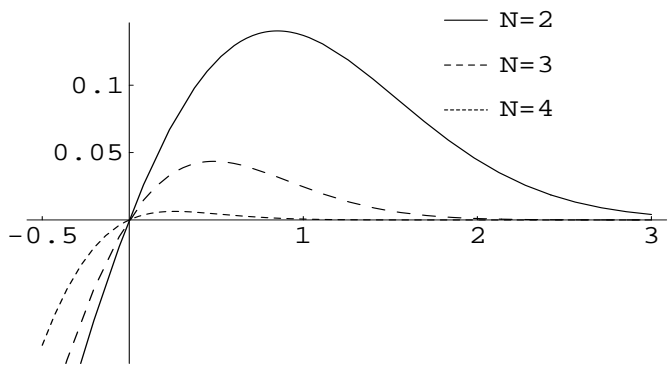}
\end{center}
\caption{The plot of the bidder's profit intensity as  a function
of deterministic withdrawal price}
\label{fig:haukcjapi}
\end{figure}
\end{center}

Let us consider in detail the case when the
player -1 fixes a unique withdrawal price, the players numbered by
$k\negthinspace=\negthinspace1,\ldots,k'\negthinspace-\negthinspace
1, k'\negthinspace+\negthinspace 1,\ldots,N$ use the same
strategies implying Gaussian distribution, but the player $k'$
unlike uses the strategy with fixed withdrawal price $e^{-q'}$
given by the Dirac measure $\delta(q_{k'}-q')dq_{k'}$. Recall
\cite{5,6} that, in the quantum approach, the logarithm of a
contingent reselling price of the good in question is indefinite
so that $E({\mathfrak p}_{k'})=0$. Therefore the $k'$-th player
profit intensity is given by
\begin{equation}
\rho_{k'}(q')=\frac{[q'\negthinspace+\negthinspace
p'\leq0]\;q'}{1+ \bigl({\int\limits^{\infty}_{q'}\eta(q)\;
dq}\bigr)^{1-N}}
\end{equation}
Fig. \ref{fig:haukcjapi} presents the shape of the profit
intensity function for the three lowest values of $N$ when
$p'\negthinspace\rightarrow\negthinspace -\infty$ and $\eta(q)$ is
the standard normal distribution.
For $N\negthinspace=\negthinspace1$ we recover the standard
$q$-bargaining of Ref. \cite{6}
and plot we be the strait line given by the equation
$\rho_{k'}\negthinspace(q'\negthinspace)=\tfrac{1}{2}q'$. Even if
there is only a few active bidders the $k'$-th player has very
limited opportunities of mak ing profits. But if she insists on
buying the good she will try to guess such a withdrawal price
$\text{e}^{p'}$ to be able to bid the possible highest price $q'$
that would not exceed $-p'$. It is worth to note here that the
quantum theory allows to multiply positive profits of a bidder
that may be meagre in a single auction. The Pauli exclusion
principle does not forbid winning in several auctions if only the
players strategy defeated the rivals (it might not result in
buying: the sellers withdrawal price might be to high). Immediate
teleportation of the state (strategy) \cite{13} makes such quantum
market technics possible and effective. The consequences of the
fact that strategies cannot be multiplied (undividity of
attention) \cite{5} resulting from the no-cloning theorem
\cite{14} are not explained by classical models. The possibility
of effective using the same strategy at different sites  allows to
make the profits arbitrary large. This paradox present in
classical approaches should incline to research into quantum
market games. The no-cloning theorem may also explain our
ignorance of our and opponents strategy states: the knowledge
would mean cloning.
\section{Conditional probabilities in quantum English auctions}The
results presented in the previous there paragraphs have to be
modified if we suppose that the players joining an auction in the
circumstances where the bidders know the prices ${\text e}^{p_k}$
at which they may resell the bought good and the price ${\text
e}^{-q_{\text -1}}$ seller paid the good  (or the value it
presents to him). Adherents of utility theory may that the
parameters  $q_{\text{-}1},p_1,\ldots,p_N$ corresponds to the
utilities of auctioned good characterizing the appropriate
players. So all participants know the value (that may depend on
the player) of the good being auctioned. In this case we should
substitute the appropriate Wigner functions \cite{5,6} for the
squared absolute values of amplitudes in $(\ref{haucja-dobicie})$:
\begin{equation}
\begin{split}
|\langle p|\psi_{\text{-}1}\rangle |^2&\longrightarrow W_{\text{-}1}(p_{\text{-}1},q_{\text{-}1})\\
|\langle q|\psi_{k}\rangle |^2&\longrightarrow W_{k}(p_k,q_k).
\end{split}
\label{haukcja-podsta}
\end{equation}
So if we take into consideration mixed strategies of participants
$\eta_{k}(p_k,q_k)$ (that is convex linear combinations of Wigner
functions) we get
\begin{equation}
\label{haucja-dobicie11}\hspace{-1.6em} dq_k \;\eta_{k}(p_k,q_k)\;
\prod_{\substack{m=1\\ m\neq k}}^{N}
\int_{-\infty}^{\infty}\negthinspace\negthinspace
dq_m\;\eta_{m}(p_m,q_m)
\int_{-\infty}^{\infty}\negthinspace\negthinspace dp_{\text{-}1}\;
\eta_{\text{-}1}(p_{\text{-}1},q_{\text{-}1})
\;[\;q_k=\negthinspace\negthinspace\min_{n=1,\ldots,N}\{q_n\}\;]\;[q_k+p_{\text{-}1}\leq0]
\end{equation}
instead of the measure $(\ref{haucja-dobicie})$. The cumulative
distribution functions
\begin{equation*}
\int_{-\infty}^p\negthinspace\eta_k(p_k,q_k\negthinspace=\negthinspace\text{constans})\;dp_k
\text{~~~~and~~~~}
\int_{-\infty}^q\negthinspace\eta_k(p_k\negthinspace=\negthinspace\text{constans},q_k)\;dq_k
\end{equation*}
have the natural interpretation of demand and supply curves of the
$k$-th player (if plotted for the common domain $\ln
c\negthinspace=\negthinspace p\negthinspace= \negthinspace-q$)
\cite{5,6,11}. The former analysis of profit intensities is now
valid only if
$p_1\negthinspace=\negthinspace\ldots\negthinspace=\negthinspace
p_N$ (except for $p_{k'}$) and if all strategies are not giffens
(the positiveness of probability measure is supposes in prove of
the theorem on maximum of profit intensity). The fascinating class
of English $q$\/-auctions with giffen strategies requires a separate
analysis.
\section{Towards a complete theory of quantum auction} The
analysis of English $q$\/-auction with reversed roles that is
bidders are selling is analogous. More interesting is the case
when the polarization of the $q$\/-auction is changed to
$|\mit1\rangle_{\text{-}1}|\mit0\rangle_1$. In this case the
player -1 reveals her withdrawal price and the player 2 accepts it
(and those of the rest of the players) or not. Such an auction is
known as the Vickrey's auction (or the second price auction). The
winner is obliged to pay the second in decreasing order price from
all the bids (and the withdrawal price of the player -1). In the
quantum approach English and Vickrey's auctions are only special
cases of a phenomenon called $q$\/-auction. In the general case
both squared absolute values of the amplitudes
$|\langle{\mit0_{\text{-}1}1_{1}}|{\mit0_{\text{-}1}1_{1}}\rangle|^2$
and
$|\langle{\mit1_{\text{-}1}0_{1}}|{\mit1_{\text{-}1}0_{1}}\rangle|^2$
are non-vanishing so we have consider them with weights
corresponding to these probabilities. Such a general $q$\/-auction
has yet no match on the existing markets. It should be very
interesting to analyse the motivation properties of $q$\/-auctions
eg finding out when the best strategy is the one corresponding to
the player's value of the good. The quantum context of the very
popular (cf the 1996 Nobel price justification) Vickrey's auction
will be analysed in a
separate paper. \\
If we consider only positive definite probability measures then
bidder gets the highest profits in Vickrey's auction using
strategies with public admission of his valuation of the auctioned
good. But it might not be so for giffen strategies because
positiveness of measures is supposed in proving incentive
character of Vickrey's auctions \cite{19}.  The presence of
giffens on real markets might not be so abstract as it seems to
be. Captain Robert Giffen who is supposed to find additive measure
not being positive definite but present on existing real market in
the forties of the XIX century \cite{16} probably got ahead of
physicists in observing quantum phenomena. Such departures from
the demand low if correctly interpreted do not cause any problem
neither for adepts nor for beginners. Employers have probably
always thought that work supply as
function of payment is scarcely monotonous. \\
The distinguished by their polarization first and second price
auctions have analogues in the Knaster solution to the pragmatic
fair division problem that is with compensatory payments for
indivisible parts of the property \cite{17}. Such a duality might
be found even in election systems that as auctions form procedures
of solving fair division problems \cite{18}. It may be that social
frustrations caused by election systems should encourage us to
discuss such topics.

\ \ \ {\bf Acknowledgments}. The authors would like to thank dr J.
Eisert for stimulating and helpful discussions.
\def\urla{\href{http://econwpa.wustl.edu:8089/eps/get/papers/9904/9904004.html}{http://econwpa.wustl.edu:8089/eps/get/papers/9904/9904004.html}}
\def\urlb{\href{http://www.spbo.unibo.it/gopher/DSEC/370.pdf}{http://www.spbo.unibo.it/gopher/DSEC/370.pdf}}
\def\urlc{\href{http://www.comdig.org}{http://www.comdig.org}}

\end{document}